\documentclass[lettersize,journal]{IEEEtran}
\usepackage{amsmath,amsfonts,amssymb}
\usepackage{array}
\usepackage{textcomp}
\usepackage{stfloats}
\usepackage{url}
\usepackage{graphicx}
\usepackage{cite}
\usepackage{algorithmic}
\usepackage{algorithm}
\usepackage{booktabs}
\usepackage{multirow}
\usepackage{helvet}
\usepackage{siunitx}
\usepackage{xcolor}
\usepackage{hyperref}
\usepackage{cleveref}
\crefname{equation}{}{}
\Crefname{equation}{Equation}{Equations}
\crefname{figure}{Fig.}{Figs.}
\crefname{table}{Table}{Tables}
\crefname{section}{Section}{Sections}
\begin{document}

\title{Reinforcement Learning to Initialize Newton-Raphson for AC Power Flow with Quantum Annealing-Based Environment Updates}

\author{Zeynab Kaseb$^{1,*}$,~\IEEEmembership{Member~IEEE,} Matthias M{\"o}ller$^{1}$, Lindsay Spoor$^3$, Jerry J. Guo$^{4,5}$, Yu Xiang$^{4,6}$, Peter Palensky$^{2}$,~\IEEEmembership{Senior~Member~IEEE,} and Pedro P. Vergara$^{2}$,~\IEEEmembership{Senior~Member~IEEE}
\thanks{
$^1${Delft Institute of Applied Mathematics, Delft University of Technology, P.O. Box 5031, 2600 GA Delft, The Netherlands.}

$^2${Electrical Sustainable Energy, Delft University of Technology, P.O. Box 5031, 2600 GA Delft, The Netherlands.}

$^3${Leiden Institute of Advanced Computer Science, Leiden University, P.O. Box 9512, 2300 RA Leiden, The Netherlands.}

$^4${Alliander N.V., P.O. Box 50, 6920 AB, Arnhem, The Netherlands.}

$^5${Intelligent Systems, Delft University of Technology, P.O. Box 5031, 2600 GA Delft, The Netherlands.}

$^6${Electrical Engineering, Eindhoven University of Technology, P.O. Box 513, 5600 MB, Eindhoven, The Netherlands.}

$^*$Corresponding author: Zeynab Kaseb (Z.Kaseb@tudelft.nl)

}}

\maketitle

\begin{abstract}
The Newton-Raphson (NR) method is widely used for solving power flow (PF) equations due to its quadratic convergence. However, its performance deteriorates under poor initialization or extreme operating scenarios, e.g., high levels of renewable energy penetration. We propose the use of reinforcement learning (RL) to optimize the initialization of NR, and introduce a quantum-enhanced RL environment update mechanism that addresses the combinatorially large action space at each RL timestep by formulating the voltage adjustment task as a Quadratic Unconstrained Binary Optimization (QUBO) problem, solved with an Ising machine. RL initialization is benchmarked against flat start and start from the DC (linearized) PF solution on a standard 4-bus system, Iwamoto's ill-conditioned 11-bus system, and the IEEE 118-bus system under normal and stressed loading and reactive power limits, with verified operational solutions. On all systems, a supervised initializer refined by RL requires fewer NR iterations than flat and DC starts and than the same initializer without RL, for all seeds. For example, on the 118-bus system under normal and stressed loading, it reached 2.04 and 2.86 NR iterations, compared with 3.02 and 5.13 from DC start and 2.61 and 3.09 without RL. In wall-clock time, this pays off only for an initializer integrated into the solver and reused for many solves on a fixed topology. On the 4-bus system, a quantum-enhanced RL agent with a quantum-inspired annealer moved challenging initial states that required 29 and 44 NR iterations to initializations that required three NR iterations within one RL timestep.
\end{abstract}


\section{Introduction}
\IEEEPARstart{T}{he} global push toward a net-zero energy future demands a paradigm shift in how electricity grids are planned, operated, and optimized. With the rapid integration of distributed energy resources (DER), electricity grids face unprecedented challenges in maintaining stability, reliability, and efficiency. At the core of this transformation lies power flow (PF) analysis, a fundamental task that enables grid operators to assess grid conditions, optimize generation dispatch, and ensure secure operation, among others~\cite{Cossent2011}.

PF analysis involves solving nonlinear algebraic equations that describe the steady state of electricity grids, which ultimately determines bus voltages, power flows, and losses. As the scale and complexity of modern electricity grids increase, addressing the computational challenges of PF analysis becomes imperative, particularly in distribution networks where the growing penetration of DERs introduces congestion. Traditional numerical techniques, e.g., the Newton-Raphson (NR) method, provide reliable solutions, but their convergence depends on the initial guess and the loading condition, and the PF equations may admit multiple solutions, of which only one corresponds to the operational state of the grid~\cite{Delabays2022}.

The NR method offers several advantages in PF analysis. Firstly, it is well established in numerical analysis and is widely supported in power system studies. Secondly, the Jacobian matrix captures the sensitivity of the power balance equations, which leads to fitting updates at each iteration. Finally, the error decreases quadratically near the solution, which, in turn, leads to quadratic convergence in most cases~\cite{Milano2009}. However, NR has several drawbacks~\cite{deOliveira2023}. Constructing the Jacobian matrix at each iteration can be computationally expensive for large-scale electricity grids. In addition, poorly chosen initial values can lead to divergence, particularly in ill-conditioned or heavily loaded electricity grids. A PF case can be categorized as ill-conditioned if, even though a physical solution exists and the grid continues to operate, it is not reachable using NR. Furthermore, the NR method may fail where the Jacobian becomes singular or nearly singular at certain operating points~\cite{Irving1987}.

In this perspective, the initial guess, $x^{0}$, significantly influences the NR method's performance~\cite{Wang2025}. A good initial guess can accelerate convergence and enhance numerical stability, while a poor initial guess may result in slow convergence or even divergence. Initial values are often chosen based on a flat start, where all voltage magnitudes are set to $1.0$~p.u. and all voltage phase angles are set to $0.0$~degrees~\cite{Alharbi2021}. The DC PF solution, which provides voltage phase angles from a linearized active power balance at unity voltage magnitudes, is another widely used starting point~\cite{Stott2009}. It is also possible to use solutions from previous timesteps in dynamic simulations as initial values~\cite{AbdelAkher2015}. Another approach is to approximate PF solutions using methods, such as fast-decoupled PF, to initialize the NR method~\cite{Coutinho2023}.

Ensuring well-chosen initial values is particularly crucial for large-scale or stressed electricity grids, where numerical instability is more likely to occur. Several studies have explored techniques to improve the initialization of the NR method. For example, integrating the NR method with gradient descent and computational graphs has been proposed to enhance convergence and computational efficiency, as well as address challenges in large-scale grids~\cite{Barati2024}. Another example is the development of parallel PF solvers, which employ high-performance computing to improve the scalability and efficiency of PF analysis in large-scale  grids~\cite{Wang2017}. Furthermore, holomorphic theory offers a non-iterative alternative that guarantees convergence to the correct operational solution, thereby mitigating issues related to ill-conditioned systems and divergence associated with the traditional NR method~\cite{Trias2012}.

Machine learning (ML) approaches have also been used to predict high-quality initial conditions from historical data, thereby improving the convergence properties of the NR method~\cite{Okhuegbe2024}. For example, convolutional neural networks have been used to provide improved initial voltage magnitude and angle estimates, which have been reported to significantly reduce solution iterations and time~\cite{Chen2020}. Techniques such as Semi-Definite Programming and Second-Order Cone relaxations have also been used to convexify the non-convex PF analysis, providing feasible solutions that can serve as effective initializations for iterative methods, such as the NR method~\cite{Molzahn2016}. Adaptive convergence enhancement strategies have also been developed to improve the robustness of PF analysis in distribution networks~\cite{Kraft1986,Yang2024}. These strategies integrate mathematical techniques, such as the superposition theorem, graph theory, and the Kron reduction method, to enhance the convergence properties of the NR method in unbalanced distribution networks.

Despite these advancements, challenges persist in selecting initial values that ensure convergence across diverse operating conditions. For example, the effectiveness of adaptive strategies can be limited by the specific characteristics of the distribution network, such as the degree of imbalance, the presence of DERs, and varying load profiles~\cite{Aviles2024}. In addition, new initialization methods are often demonstrated on selected cases, and systematic comparisons and acceptance criteria remain scarce. In particular, it is rarely examined whether NR actually fails under realistic stress, whether a converged solution is the operational one, and how generator reactive power limits, which convert voltage-controlled (PV) buses into load (PQ) buses during the solution process~\cite{Zimmerman2011}, affect convergence~\cite{Zeng2022,Song2024}.

Reinforcement learning (RL)~\cite{Sutton2018} is a subfield of ML that enables an agent to learn through trial and error and make sequential decisions by interacting with an environment. In electricity grid applications, RL has emerged as a promising approach for addressing complex control problems, such as voltage regulation and optimal PF, especially in the presence of uncertainties and nonlinearities~\cite{Ma2024,Wang2025b}. In this regard, RL agents can adapt their decisions to the operating conditions using learned system-specific patterns and past experiences~\cite{Yang2020}, which motivates their use for NR initialization. We therefore propose an RL-based NR initialization that helps steer the NR method toward regions of the solution space with a higher likelihood of convergence. In doing so, the agent must determine an action at each RL timestep, specifically, choosing increments to adjust the voltage magnitudes and angles across all buses. The resulting state of the power system is then evaluated for a single combination of these adjustments per RL timestep. Nevertheless, due to the combinatorial nature of voltage modifications, classical solvers can become computationally inefficient when evaluating state transitions~\cite{Chicano2025}.

Quantum computing (QC) is an emerging technology that enables new possibilities in simulation, combinatorial optimization, convex optimization, and ML approaches, among others~\cite{Morstyn2024}. Adiabatic Quantum Computing (AQC), particularly through quantum annealing, offers a promising approach for addressing the combinatorial complexity of RL-based initialization. AQC utilizes the adiabatic theorem to evolve a quantum system toward the ground state of a problem-specific Hamiltonian, effectively encoding and solving complex optimization problems. AQC is computationally equivalent to the standard gate-based QC model~\cite{Aharonov2007}, and digitized adiabatic evolution has been demonstrated on superconducting hardware~\cite{Barends2016}. By formulating the voltage adjustment task as a Quadratic Unconstrained Binary Optimization (QUBO) problem, AQC can, in principle, explore the high-dimensional action space to identify optimal or near-optimal voltage updates~\cite{Kaseb2025solving,Morstyn2024}. Therefore, this paper also proposes a novel quantum-enhanced RL environment update mechanism to strategically guide initialization.

Three test systems are used, each for a specific purpose. A standard 4-bus system is small enough to map the basin of attraction of NR exhaustively and to solve the QUBO on currently available quantum(-inspired) hardware. Iwamoto's 11-bus system is a classical ill-conditioned case for which NR from conventional starting points is known to fail~\cite{Iwamoto1981}. The IEEE 118-bus system, with $53$ PV buses, tests RL in a high-dimensional action space and PV$\to$PQ switching under reactive power limits. RL is compared with the flat start and start from the DC (linearized) PF solution under normal and stressed loading, with an independent acceptance test of the operational PF solution. The QUBO of the quantum-enhanced RL (QRL) environment update is solved on D-Wave's Advantage\texttrademark\ system (QA) and with Fujitsu's Quantum-Inspired Integrated Optimization software (QIIO). The main contributions of this work are:
\begin{itemize}
\item An initializer refined by RL that requires fewer NR iterations than the flat and DC starts on all three systems. We also identify when its training and evaluation costs pay off.
\item Incorporation of quantum and quantum-inspired annealers into the RL environment update mechanism to obtain the state of the power system based on a problem-specific Hamiltonian formulated for PF analysis, searching all combinations of voltage increments heuristically in a single RL timestep.
\item A quantitative benchmark of RL-based initialization against flat and DC starts on test systems up to the 118-bus system, including an ill-conditioned system, under stressed loading and generator reactive power limits with PV$\to$PQ switching.
\end{itemize}

\section{Power Flow Analysis}\label{sec:powerflow}

PF analysis aims to compute the unknown parameters for different bus categories in an electricity grid that satisfy power balance equations, which can be compactly expressed as a root-finding problem:
\begin{equation} \label{eq:compact-pf}
    F(x) : \quad \mathbf{S} -  \mathbf{V} \circ (\mathbf{Y}\mathbf{V})^* = 0 \,
\end{equation}
where $x$ denotes the state variables, which include unknown parameters, e.g., the complex voltages at \emph{load} buses.

Note that \cref{eq:compact-pf} is nonlinear and non-convex, and hence, there is no analytical solution to it. Therefore, PF analysis is generally performed using iterative numerical methods, such as the NR and Gauss-Seidel methods. These methods start with initial values for unknown parameters $x^{(0)}$ at $\text{it} = 0$. The iterative procedure to update $x^{(\text{it}+1)}$ can be written as:
\begin{equation} \label{eq:numerical-solution}
    x^{(\text{it}+1)} = x^{(\text{it})} + w^{(\text{it})} \Delta x{,}
\end{equation}
where $w^{(\text{it})}$ is typically a relaxation factor that controls how much of the computed update $\Delta x$ is applied at each iteration.

NR is one of the most widely used iterative methods for PF analysis due to its quadratic convergence properties and robustness in well-conditioned cases~\cite{Tinney1967}. The method is based on a first-order Taylor series expansion of the PF equations around an operating point. The update step $\Delta x$ is computed, as:
\begin{equation} \label{eq:NR-step}
\Delta x = - J^{-1} (x^{(\text{it})}) F(x^{(\text{it})}),
\end{equation}
where $J(x) = \partial F(x)/\partial x$ is the Jacobian matrix evaluated at the current iteration $x^{(\text{it})}$.

At each iteration, the Jacobian matrix is computed and used to solve for $\Delta x$, which is then applied to update the state variables $x$, as calculated in \cref{eq:numerical-solution}. The process continues until convergence criteria, such as $\max|F(x)| \leq \epsilon$, are met, where $\epsilon$ is a predefined tolerance.

$F$ collects the active power mismatches at PV and \emph{load} buses and the reactive power mismatches at \emph{load} buses, and NR stops when $\max|F(x)| \leq \epsilon = 10^{-8}$~p.u.\ within at most $50$ NR iterations, as implemented in the PandaPower Python package~\cite{Thurner2018}. Since the PF equations may admit multiple solutions~\cite{Delabays2022}, a converged NR run is accepted only if its independently recomputed mismatch meets the same tolerance, PV and \emph{slack} voltage magnitudes equal their set-points, and the solution coincides with the operational solution within $10^{-4}$~p.u. The operational solution is obtained by warm-started NR along increasing load from $20\,\%$ loading, i.e., a simple load continuation~\cite{Ajjarapu1992}.

A flat start sets all voltage magnitudes to $1.0$~p.u.\ (set-points at PV and \emph{slack} buses) and all angles to the \emph{slack} angle. A DC start takes the voltage angles from the DC PF solution~\cite{Stott2009} and requires one additional linear solve of the DC PF equations, which is not counted in the NR iterations in this work. When the reactive power of a generator violates its limits, the corresponding PV bus is converted into a PQ bus with its reactive power fixed at the violated limit, and NR is repeated in an outer loop, as in MATPOWER~\cite{Zimmerman2011} and PandaPower~\cite{Thurner2018}. Such switching, here one-directional (PV$\to$PQ), can cause numerical difficulties and multiple solutions~\cite{Zeng2022,Song2024}.

\section{Markov Decision Process (MDP)}

MDP is the typical mathematical formulation for a standard RL framework, defined by the tuple~\cite{Sutton2018}:
\begin{equation}
    \mathcal{M} = \langle S, A, R, P, \gamma, d_0 \rangle,
\end{equation}
where $S$ is the set of states representing all possible configurations of the environment, $A$ is the set of actions available to the agent, $R: S \times A \times S \to \mathbb{R}$ is the reward function that provides scalar feedback for taking action $a$ in state $s$, $P: S \times A \times S \to [0,1]$ defines the transition dynamics of the environment, i.e., a probability distribution over all next states given a state $s$ and action $a$, $\gamma\in (0,1)$ is the discount factor that balances the trade-off between rewards of immediate and future timesteps, and $d_0$ is the initial state distribution over $S$.

In an MDP, the transition dynamics distribution $P$ is Markovian, meaning that the next state given an action only depends on the current state. At each timestep during training, the RL agent observes the state of the environment, performs an action, and the environment transitions to the next state. The RL agent then receives feedback as a reward for taking that action, which guides the learning process to optimize a long-term objective that is accelerating NR convergence.

Through interaction with the environment on a diverse set of system conditions, the RL agent learns a decision-making strategy, called a policy $\pi: S \to A, s \mapsto \pi(a|s)$, which defines a probability distribution over actions given a state. The main objective is to find a policy that maximizes the expected cumulative discounted reward over time:
\begin{equation} \label{eq: RL_objective}
    J(\pi) = \mathbb{E}_{\tau \sim \pi}\left[\sum_{t=0}^{\infty}\gamma^t R(s_t, a_t) \right],
\end{equation}
where $\tau=\{s_0,a_0, \dots s_t, a_t, \dots\}$ $\forall s \in S, a \in A$ is defined as a trajectory sampled from policy $\pi$. Note that throughout this paper, a conventional notation for the reward at timestep $t$ is used, denoted as $r_t$, which is equivalent to $R(s_t, a_t)$ in \cref{eq: RL_objective}. The optimal policy can then be found by maximizing the objective over the entire set of policies:
\begin{equation}
    \pi^{*} = \arg \max_{\pi} J(\pi).
\end{equation}

To achieve an optimal policy, one can define state-value and state-action-value functions for a policy $\pi$, $V_{\pi}(s)$ and $Q_{\pi}(s,a)$, which estimate the expected future reward starting from a given state $s$ or state-action pair $(s,a)$, respectively:
\begin{equation} \label{eq: Bellman_V}
    V_{\pi}(s) = \mathbb{E}_{a \sim \pi, s' \sim P} \left[ R(s, a) + \gamma V_\pi(s') \right].
\end{equation}
\begin{equation} \label{eq: Bellman_Q}
    Q_\pi(s,a) = \mathbb{E}_{s' \sim P} \left[ R(s, a) + \gamma \mathbb{E}_{a' \sim \pi} Q_\pi(s', a') \right].
\end{equation}

\cref{eq: Bellman_V,eq: Bellman_Q} are referred to as the \textit{Bellman Equations} for $V_{\pi}$ and $Q_{\pi}$, respectively. However, these equations assume that the transition dynamics $P$ are known, which oftentimes is not the case in RL. Therefore, most RL methods rely on an agent collecting experiences, i.e., trajectories $\tau \sim \pi$, by interacting with the environment. Leveraging these experiences, the value functions $V_{\pi}$ and $Q_{\pi}$ can be learned with \textit{value-based} methods. Alternatively, a policy can also be directly learned, which is done by \textit{policy-based} methods. Current state-of-the-art RL algorithms learn both the value functions and a policy, which is referred to as \textit{actor-critic}~\cite{Konda1999}.

\section{Proximal Policy Optimization (PPO)}

PPO is a widely used actor-critic algorithm in modern deep RL that aims to improve learning stability and efficiency by constraining policy updates while maximizing cumulative rewards~\cite{Schulman2017}. Here, deep RL refers to approaches that use neural networks (NNs) parametrized by a set of weights to approximate value functions or policies. PPO combines the strengths of policy-gradient methods with clipped surrogate objectives to prevent large, destabilizing updates, thus offering a practical balance between exploration and exploitation. The PPO model consists of the following components:
\begin{itemize}
    \item \textbf{Policy Network} is a parameterized policy $\pi_\phi(a_t | s_t)$, represented by an NN, where $\phi$ refers to the NN parameters, $s_t$ represents the state at timestep $t$, and $a_t$ is the action to be taken at that state. In the context of PF analysis, the policy network takes the current state $s_t$ (\cref{sec:setup}) as input and outputs the parameters of a probability distribution (e.g., mean and standard deviation for Gaussian policies in continuous action spaces). During training, PPO updates the policy network using stochastic gradient ascent on a clipped surrogate objective, preventing policy updates from being too large. The direction and magnitude of each update are guided by an advantage function, which quantifies the improvement of an action over the average action under the current policy.

    \item \textbf{Value Network} estimates the expected return from a given state under the current policy. It provides a baseline for computing the advantage function. In mathematical terms, the value function \( V^\pi_{\phi'}(s) \) is defined as:
    \begin{equation}
    V^\pi_{\phi'}(s) = \mathbb{E}_\pi \left[ \sum_{t=0}^{T} \gamma^t r_t \mid s_0 = s \right]
    \end{equation}
    where $\phi'$ are the parameters of the parametrized NN that approximates the value function and $T$ is the episode length. This network is trained using a value loss, such as the mean squared error between predicted and actual returns, and reduces the variance in policy updates.

    \item \textbf{Learning Procedure} refers to the iterative interactions with the environment, as outlined in \cref{alg:RL_learning}, through which the PPO agent improves its policy. The policy and value networks are trained jointly using mini-batch gradient updates over trajectories collected during agent-environment interaction. The PPO loss function includes both the clipped surrogate objective for the policy and a value loss term, optionally regularized with an entropy bonus to encourage exploration. The policy update of an epoch can be stopped early when the approximate Kullback--Leibler (KL) divergence between the old and the new policy exceeds a target value~\cite{Raffin2021}, which prevents destabilizing updates in high-dimensional action spaces.
\end{itemize}

\begin{algorithm}[t]
\caption{Learning procedure of the RL agent.} \label{alg:RL_learning}
\small
\begin{algorithmic}[1]
    \STATE \textbf{Initialize} $\pi_\phi$ and $V^\pi_{\phi'}$
    \WHILE{training}
    \STATE Collect experience in the environment using $\pi_\phi$
    \STATE Compute advantage estimates $A_t$
    \renewcommand{\algorithmicfor}{\textbf{for}}\renewcommand{\algorithmicdo}{\textbf{do}}%
    \renewcommand{\algorithmicendfor}{\textbf{end for}}%
    \FOR{each epoch and mini-batch}
    \STATE Update $\pi_\phi$ with the clipped surrogate objective
    \STATE Update $V^\pi_{\phi'}$ with the value loss
    \ENDFOR
    \STATE Stop early if approximate KL divergence $>$ target
    \ENDWHILE
\end{algorithmic}
\end{algorithm}

\section{Experimental Setup} \label{sec:setup}

\subsection{Test systems and operating scenarios} \label{sec:scenarios}
Three test systems are considered (\cref{tab:systems}): the standard 4-bus system~\cite{Grainger1994}, Iwamoto's ill-conditioned 11-bus system~\cite{Iwamoto1981}, and the IEEE 118-bus system, as provided by PandaPower~\cite{Thurner2018}. A load scenario scales every load at constant power factor by $\lambda\,\xi_l$, with $\xi_l \sim \mathcal{U}(1-\sigma, 1+\sigma)$, and every generator active power by $\lambda$, while the \emph{slack} bus balances the system. Under \emph{normal} loading, $\sigma = 0.20$ and $\lambda \in [0.70, 1.10]$ times the nominal load for the 4- and 118-bus systems and $\lambda \in [0.70, 0.90]$ for the 11-bus system, whose nominal load lies at the continuation estimate. Under \emph{stressed} loading, $\lambda \in [0.90, 0.97]\,\hat{\lambda}$ and $\sigma = 0.05$, where $\hat{\lambda}$ is the largest uniform loading reached by the load continuation of \cref{sec:powerflow}. $\hat{\lambda}$ is an estimate and not a certified loadability limit. Reactive power limits are ignored at these two levels (a numerical load-scaling benchmark) and studied in \cref{sec:qlim} relative to the estimate $\hat{\lambda}_Q$ with limits enforced. For every system and level, $300$, $100$, and $200$ scenarios form disjoint training, validation, and test sets, respectively. The stress levels were fixed on separate pilot scenarios before any test evaluation.

\begin{table}[t]
\centering
\caption{Test systems, continuation-based loading estimates without ($\hat{\lambda}$) and with ($\hat{\lambda}_Q$) generator reactive power limits, and the stressed loading range.}
\label{tab:systems}
\footnotesize
\setlength{\tabcolsep}{4pt}
\begin{tabular}{lcccccc}
\toprule
System & Buses & PV & PQ & $\hat{\lambda}$ & $\hat{\lambda}_Q$ & Stressed $\lambda$ \\
\midrule
4-bus & 4 & 0 & 3 & 2.41 & -- & 2.17--2.34 \\
11-bus & 11 & 0 & 10 & 1.00 & -- & 0.90--0.97 \\
118-bus & 118 & 53 & 64 & 3.19 & 2.14 & 2.87--3.09 \\
\bottomrule
\end{tabular}
\end{table}

\subsection{Baselines and performance metrics}
RL-based initialization is compared with flat and DC starts, all solved using the same NR implementation and evaluated using the acceptance test in \cref{sec:powerflow}. The reported metrics are the success rate, i.e., the share of test scenarios whose NR run is accepted, and the mean number of NR iterations over accepted runs. A lower iteration count is only regarded as an improvement at an equal success rate. Two RL-based initializers are evaluated (\cref{sec:mdp}): \emph{RL}, the sequential formulation, and \emph{RL-refined}, a supervised initializer (\emph{supervised}) refined by RL. For RL-refined, all three seeds are reported (mean $\pm$, standard deviation std), each compared with its own supervised initializer. For RL, the best of three seeds on validation data is reported. The NR iteration count measures the computational effort. Wall-clock times are analyzed in \cref{sec:cost}.

\subsection{RL formulation} \label{sec:mdp}
\Cref{alg:episode} and \cref{fig:diagram} depict an RL episode for selecting complex voltage adjustments following a policy. An episode in RL refers to one complete run of the agent in the environment, from the initial state until the agent reaches a terminal state or until the maximum episode length $T$ is reached. Each step refers to one interaction between the agent and the environment. At each step, the agent selects an action. Next, the environment processes that action, and transitions into the next state. Then, the agent receives a new state and reward. Over the course of an episode, the agent takes multiple timesteps to reach a solution. The MDP is characterized as:

\begin{algorithm}[t]
\caption{RL episode for selecting complex voltage adjustments following a policy $\pi$.} \label{alg:episode}
\small
\begin{algorithmic}[1]
\STATE \textbf{Initialize} Environment from a scenario, policy $\pi$, $T$, $t \gets 0$
\STATE Get initial state $s_0 = [\vec{V}^0, \vec{\theta}^0, {\log_{10}\max|F(x^0)|, \vec{P}, \vec{Q}}]$
\WHILE{$t < T$}
    \STATE Based on the current state $s_t$, the agent selects action:
    \STATE \hspace{0.55cm}$a_t \sim \pi(\cdot \mid s_t)$
    \STATE Perform environment update using action $a_t$
    \STATE Get new state $s_{t+1}$, reward $r_{t+1} = -\|F(x^{t+1})\|_2$
    \IF{$\max|F(x^{t+1})| \leq 10^{-5}$~p.u.}
        \STATE Episode done; \textbf{break}
    \ENDIF
    \STATE $t \gets t+1$, $s_t \gets s_{t+1}$
\ENDWHILE
\STATE Episode terminated; initialize NR with the final state
\end{algorithmic}
\end{algorithm}

\begin{itemize}
    \item The initial state distribution $d_0$ consists of the flat start of a load scenario drawn from the training set, so that the learned initialization is compared with flat and DC starts for the same operating condition.
    \item The state $s_t$ at timestep $t$ contains the voltage magnitudes $\vec{V}^t$ and phase angles $\vec{\theta}^t$ (relative to the \emph{slack} bus) at non-\emph{slack} buses, the PF mismatch $\log_{10}\max|F(x^t)|$, and the active and reactive power injections of the scenario. The number of NR iterations is not part of the state.
    \item The actions consist of normalized increments $a_t \in [-1,1]^{d}$. For the QRL comparison on the 4-bus system, the actions are the complex voltage adjustments $\Delta \mu_i^t \in (-0.5,0.5]$~p.u.\ and $\Delta \omega_i^t \in (-0.25,0.25]$~p.u.\ in rectangular coordinates at \emph{load} bus $i$, which are compatible with the QUBO update of \cref{sec:qrl}. For the comparison across systems, a PF-compatible action space is used, i.e., angle increments of up to $\pm 0.25$~rad at PV and \emph{load} buses and voltage magnitude increments of up to $\pm 0.5$~p.u.\ at \emph{load} buses, while the \emph{slack} bus and the PV voltage magnitudes remain at their set-points.
    \item The reward is defined by a penalizing term based on the norm of the residual vector $F(x^t)$, where $\Delta P_i^t$ and $\Delta Q_i^t$ are the active and reactive power mismatches resulting from applying the updated initial values in PF analysis, $r_t = -\|F(x^t)\|_2$. Unlike the NR iteration count, which saturates at $50$ for $26$--$100\,\%$ of random first steps, the mismatch norm distinguishes all states.
    \item An episode terminates when $\max|F(x^t)| \leq 10^{-5}$~p.u.\ or after $T = 20$ timesteps. The final state is then used to initialize NR with the tolerance of \cref{sec:powerflow}.
\end{itemize}

For the QRL results of \cref{sec:qrl}, the MDP draws the initial voltages at \emph{load} buses uniformly from $(0, 2]$~p.u.\ and $(-90, 90]$~degrees at nominal load, includes the NR iteration count in the state, and uses $[32, 32, 32]$ neurons.

To optimize the policy, we employ PPO~\cite{Schulman2017} using the implementation from Stable-Baselines3~\cite{Raffin2021}, and train the RL agent for $\num{1e5}$ timesteps. The PPO settings are learning rate $0.7 \times 10^{-4}$, discount factor $0.9$, $2048$ steps per update, batch size $64$, $10$ epochs, clipping range $0.2$, no entropy bonus, policy and value networks with $[64, 64, 64]$ neurons and rectified linear unit (ReLU) activation, initial policy standard deviation $0.1/\sqrt{d}$, and, for the agents with the PF-compatible action space, a target KL divergence of $0.02$. Observations and rewards are normalized with running statistics frozen after training. The episode length, learning rate, initial standard deviation, and reward were tuned on validation data. Per agent, the checkpoint with the lowest mean validation NR iteration count (failures counted as $50$) is selected.

For RL refinement of a supervised initializer, the corrections from the DC start to the operational solutions of the training scenarios (angles relative to the \emph{slack} bus and voltage magnitudes at \emph{load} buses, scaled by $0.1$~rad and $0.05$~p.u.) are approximated by their mean plus a weighted sum of their eight leading principal components (all six for the 4-bus system; $99.998\,\%$ of the variance for the 118-bus system). An NN with two hidden layers of $64$ neurons predicts the eight weights, i.e., the coefficients, from the power injections of the scenario. The RL agent observes the injections, the mismatch vector at this supervised initialization, and the predicted coefficients, and then applies a single bounded correction to the coefficients. The reward is the verified reduction in NR iterations relative to the supervised initializer (failures counted as $50$), with a small mismatch term that only breaks ties. PPO ($[64, 64]$ neurons, learning rate $10^{-4}$, $\gamma = 1$, target KL divergence $0.02$) is trained for $\num{20480}$ episodes with three seeds, and checkpoints are selected on validation data.

\section{Results}

\subsection{Basin of attraction of NR} \label{sec:4bus}

\begin{figure}[t]
\centering
\includegraphics[width=\columnwidth]{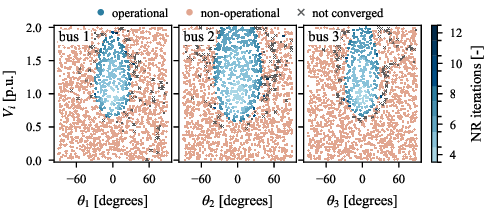}
\caption{Basin of attraction for initial values at \emph{load} buses for the 4-bus test system at nominal load. Each subplot shows the number of NR iterations (indicated by the heatmap) required to perform PF analysis for the given voltage magnitude (p.u.) and phase angle (degrees), starting from initial values that NR converges to operational solutions. Orange points converge to a non-operational/low-voltage solution. Crosses do not converge within $50$ iterations.}
\label{fig:distribution}
\end{figure}

For a standard 4-bus test system~\cite{Grainger1994}, consisting of one \emph{slack} bus with voltage $1.0\angle0.0$~p.u.\ and three \emph{load} buses with nominal active and reactive power demands $\vec{P}^\text{D} = [1.70, 2.00, 0.80]$ and $\vec{Q}^\text{D} = [1.054, 1.239, 0.496]$~p.u.\ ($100$~MVA base), the basin of attraction and regions of initial voltage magnitudes (p.u.) and voltage phase angles (degrees) at \emph{load} buses are illustrated in \cref{fig:distribution}. For each subplot, the initial values of the corresponding bus are varied while the initial values for the other buses are kept fixed at $1\angle0$~p.u. Only $15.6$--$27.4\,\%$ of the $2000$ random initial points per bus converge to the operational solution, whereas $67.9$--$80.4\,\%$ converge to a low-voltage solution of the PF equations and $3.6$--$4.8\,\%$ do not converge within $50$ iterations. From a flat start, NR converges to the operational solution in four iterations. Accordingly, improper initialization of complex voltages in certain regions can lead to slow convergence of the NR method, to divergence, or to a physically meaningless solution. To mitigate this risk, we train an RL agent to adjust the initial complex voltage estimates, steering them toward regions that ensure NR convergence while minimizing the number of iterations required for PF analysis.

\subsection{Convergence of flat and DC starts under stress} \label{sec:scaling}
\begin{figure}[t]
\centering
\includegraphics[width=\columnwidth]{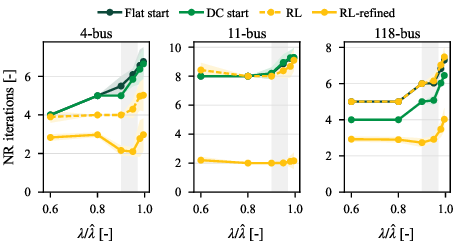}
\caption{NR iterations (mean $\pm$ std over $40$ pilot scenarios per point) versus loading relative to $\hat{\lambda}$, reactive power limits ignored: flat and DC starts, RL, and RL-refined (\cref{sec:latent_res}; mean over three seeds). Gray: stressed range. All runs converged to the operational solution. The DC start requires one additional (linear) DC PF iteration, which is not included.}
\label{fig:stress}
\end{figure}

Before evaluating RL, it is examined whether NR convergence is genuinely problematic. For Iwamoto's 11-bus system, NR fails at nominal load from flat and DC starts~\cite{Iwamoto1981}, but the load continuation also stops at $0.998$ times the nominal load, i.e., no method found a solution. On pilot scenarios with loading up to $0.995\,\hat{\lambda}$ (\cref{fig:stress}), NR from flat and DC starts converged to the operational solution in all $720$ scenarios per start, and the mean iteration count increased from $4$--$5$ at $0.6\,\hat{\lambda}$ to $6.5$--$7.3$ at $0.995\,\hat{\lambda}$ for the 4- and 118-bus systems and from $8.0$ to $9.3$ for the 11-bus system. With reactive power limits on the 118-bus system, two pilot scenarios close to $\hat{\lambda}_Q$ failed for every start alike, i.e., due to PV$\to$PQ switching rather than the initialization, and load steps of $15$--$35\,\%$ and single line outages caused no failures in $36$ pilot cases. Hence, stress increases the number of NR iterations rather than causing divergence, and RL is evaluated for reducing NR iterations at equal reliability.

\subsection{RL and RL-refined initialization} \label{sec:latent_res}

\begin{figure}[t]
\centering
\includegraphics[width=\columnwidth]{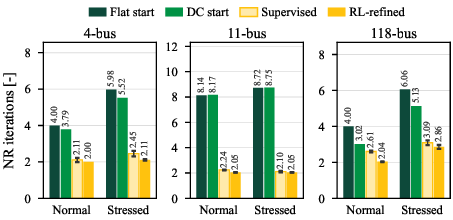}
\caption{Mean NR iterations on the test sets ($200$ scenarios per system and level) from flat and DC starts, supervised, and RL-refined (mean over three seeds; error bars show std over seeds). All runs converged to the operational solution. The DC start requires one additional (linear) DC PF iteration, which is not included.}
\label{fig:iterations}
\end{figure}

RL requires $3.57$ and $4.17$ NR iterations for the 4-bus system under normal and stressed loading, respectively, which is $0.22$ and $1.35$ fewer than the DC start, and $7.94$ and $8.19$ for the 11-bus system, i.e., $0.23$ and $0.56$ fewer. \Cref{fig:iterations} shows the test results for flat and DC starts as well as supervised and RL-refined. On the 118-bus system, the supervised initializer reduces the NR iterations from $3.02$ and $5.13$ for DC start to $2.61 \pm 0.05$ and $3.09 \pm 0.13$ under normal and stressed loading, and RL-refined further to $2.04 \pm 0.02$ and $2.86 \pm 0.11$ (mean $\pm$ std over seeds). On the 4- and 11-bus systems, RL-refined requires $2.00$ and $2.11$, and $2.05$ and $2.05$ NR iterations, compared with $3.79$ and $5.52$, and $8.17$ and $8.75$ from DC start, and $2.11$ and $2.45$, and $2.24$ and $2.10$ without RL. The contribution of RL is positive for all nine agents and both levels, i.e., $0.53$--$0.62$ and $0.16$--$0.30$ NR iterations on the 118-bus system and $0.01$--$0.46$ on the smaller systems, where the supervised initializer already requires close to two iterations. On the pilot scenarios of \cref{fig:stress}, which include loadings outside the training range, RL-refined requires $2.0$--$3.0$, $2.0$--$2.2$, and $2.7$--$4.0$ NR iterations on the 4-, 11-, and 118-bus systems, respectively. All runs reached the operational solution.

\subsection{Computational cost} \label{sec:cost}
All times were measured in a single process on an Intel Core i7-1185G7 CPU (4 cores, 3.0~GHz) with 16~GB RAM under Windows~11, using Python~3.11, PandaPower~2.14 with Numba, and PyTorch~2.14 on the CPU. For the 118-bus system, the median time per scenario (initialization and NR solve in PandaPower, $60$ validation scenarios, three repetitions) is $68$ and $72$~ms from PandaPower's built-in DC start, the relevant practical alternative, and $101$ and $107$~ms from a DC start computed explicitly as for the learned initializers, compared with $102$ and $104$~ms for the supervised initializer and $107$ and $111$~ms for RL-refined (normal and stressed). The explicit path builds the network model twice and solves the DC problem separately ($33$--$35$~ms), whereas the learned components require only $0.38$~ms (supervised prediction) and $0.27$~ms (mismatch and policy), and one NR iteration costs about $2.5$~ms. The offline cost per seed is about $14$~min, that is, $229$~s for the operational solutions of the $600$ training scenarios, $30$~s for DC starts and the supervised fit, and $578$~s for RL training.

Hence, the additional training, preprocessing, and policy evaluation are worthwhile only if (i) the initializer is integrated into the solver, reusing the network model and DC solve of the built-in DC start. Then, the $0.98$ and $2.27$ iterations saved relative to the DC start ($2.5$ and $5.7$~ms) exceed the $0.65$~ms of the learned components, and the RL step alone saves $1.4$ and $0.6$~ms at a cost of $0.27$~ms, whereas without integration, the built-in DC start is faster by about $39$~ms; (ii) NR iterations are expensive relative to the learned components, i.e., for large systems, where the sparse Jacobian factorization grows faster with the system size than the NN evaluation, and under stressed or ill-conditioned conditions, where more iterations are saved, e.g., six on the 11-bus system; and (iii) many PF problems are solved for a fixed topology and bus types, e.g., in time series or Monte Carlo studies, so that the offline cost is amortized, i.e., after about $4.6 \times 10^5$ and $1.7 \times 10^5$ solves at net savings of $1.8$ and $5.0$~ms. For single solves, changing topologies, or small systems, the DC start remains preferable.

\subsection{Quantum-enhanced RL environment update mechanism} \label{sec:qrl}

\begin{figure}[t]
\centering
\includegraphics[width=0.8\columnwidth]{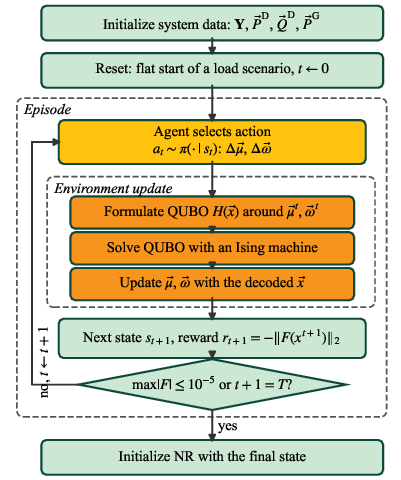}
\caption{Schematic of the quantum-enhanced RL update mechanism for PF analysis based on the QUBO representation. The diagram depicts the episode structure, state transitions, and quantum-enhanced RL environment updates. Key functions, i.e., reset, action, state, and reward, govern the agent's interaction, while the Ising machine selects the voltage adjustments. For classical RL, the environment update applies the action directly.}
\label{fig:diagram}
\end{figure}

A remaining challenge is to update the state $s$ more efficiently based on the agent's selected actions $a$. Given the combinatorial nature of possible complex voltage adjustments, we introduce a quantum-enhanced RL environment update mechanism. Specifically, the environment update is based on a problem Hamiltonian defined for a combinatorial reformulation of PF equations. We then use Ising machines to determine voltage updates within the environment. \Cref{fig:diagram} shows the episode structure, state transitions, and quantum-enhanced RL environment updates. The agent interacts with the environment through key functions, including reset, action, state, and reward, while the Ising machine selects the complex voltage adjustments.

The pseudo-code of the proposed update mechanism, i.e., the \emph{Environment Update} block in \cref{fig:diagram}, can be found in our previous study~\cite{kaseb2024power}. Accordingly, the power system data is first initialized, including load power demands, $\vec{P}^\text{D}$ and $\vec{Q}^\text{D}$, generation power outputs $\vec{P}^\text{G}$, and the admittance matrix $\mathbf{Y}$. The RL agent then provides the action variables, which correspond to complex voltage adjustments in rectangular coordinates, $\vec{\Delta{\mu}}$ and $\vec{\Delta{\omega}}$. The voltage magnitudes $\vec{V}$ and phase angles $\vec{\theta}$ at \emph{load} buses are extracted from the state \(s_t\) and converted to rectangular coordinates, yielding $\mu = V \cos(\theta)$ and $\omega = V \sin(\theta)$.

The base values of $\vec{\mu}^0$ and $\vec{\omega}^0$ are stored. A problem Hamiltonian, $H(\vec{x})$, is then formulated using a QUBO representation. This Hamiltonian encodes the combinatorial PF formulation, whose low-energy states correspond to state updates with a small PF mismatch. In doing so, PF equations can be expressed as separated active and reactive components, which yields:
\begin{subequations} \label{eq:pq-balance}
    \begin{align}
        \vec{P} - \vec{P}^\text{G} + \vec{P}^\text{D} & = 0, \label{eq:p-balance}\\
        \vec{Q}- \vec{Q}^\text{G} + \vec{Q}^\text{D} & = 0, \label{eq:q-balance}
    \end{align}
\end{subequations}
where $\vec{P}$ is the vector of net active power injections and $\vec{Q}$ is the vector of net reactive power injections. The vectors of generated active power $\vec{P}^\text{G}$, generated reactive power $\vec{Q}^\text{G}$, consumed active power $\vec{P}^\text{D}$, and consumed reactive power $\vec{Q}^\text{D}$ are assumed to be known. To convert the problem into a combinatorial optimization problem suitable for Ising machines, we express the complex voltages in rectangular coordinates:
\begin{subequations} \label{eq:pq-expanded}
    \begin{align}
        P_i & = \sum_{k=1}^{N} \mu_i G_{ik} \mu_k + \omega_i G_{ik} \omega_k + \omega_i B_{ik} \mu_k - \mu_i B_{ik} \omega_k,  \label{eq:p-expanded}\\
        Q_i & = \sum_{k=1}^{N} \omega_i G_{ik} \mu_k - \mu_i G_{ik} \omega_k - \mu_i B_{ik} \mu_k - \omega_i B_{ik} \omega_k, \label{eq:q-expanded}
    \end{align}
\end{subequations}
where $P_i$ and $Q_i$ are the net active power and the net reactive power at bus $i$, respectively.

Here, we discretize $\mu_i$ and $\omega_i$ to accommodate binary decision variables in the formulations:
\begin{subequations} \label{eq:muomega-increment}
    \begin{align}
        \mu_i & = \mu_i^0 + \Delta \mu_i (x_{i,0}^\mu - x_{i,1}^\mu), \label{eq:mu-increment}\\
        \omega_i & = \omega_i^0 + \Delta \omega_i (x_{i,0}^\omega - x_{i,1}^\omega), \label{eq:omega-increment}
    \end{align}
\end{subequations}
where $x_{i,\{0,1\}}^{\{\mu,\omega\}}\in\{0,1\}$ are binary decision variables indicating whether the base values $\mu_i^0$ and $\omega_i^0$ are increased, decreased, or left unchanged. Replacing \(\mu_i\) and \(\omega_i\) in \cref{eq:pq-expanded} with \cref{eq:muomega-increment} yields extended formulations for $P_i$ and $Q_i$, respectively. Detailed information about the QUBO representation can be found in~\cite{kaseb2024power}. To solve \cref{eq:pq-balance} using Ising machines, the objective is to minimize the squared sum of all terms, expressed as:
\begin{equation} \label{eq:hamiltonian}
\begin{aligned}
    &\min_{x \in \{0,1\}^{4(N-1)}} H(\vec{x}) \quad \text{with} \\
    &H(\vec{x}) = {\sum_{i \in \mathcal{N}_P}} (P_i - P_i^\text{G} + P_i^\text{D})^2 + {\sum_{i \in \mathcal{N}_Q}} (Q_i - Q_i^\text{G} + Q_i^\text{D})^2{,}
\end{aligned}
\end{equation}
where $\mathcal{N}_P$ contains the PV and \emph{load} buses and $\mathcal{N}_Q$ the \emph{load} buses, i.e., only the specified balances enter, whereas the \emph{slack} injections and the reactive power of PV buses are free. For the 4-bus system, both contain the three \emph{load} buses.

In this work, we utilize two Ising machines, i.e., QA and QIIO with a predefined number of reads $n_{\text{read}}$. Note that expanding the terms in \cref{eq:hamiltonian} results in a fourth-order polynomial in binary variables. Both Ising machines accept quadratic terms only. We use the Python package PyQUBO\footnote{\url{https://pyqubo.readthedocs.io}} and Fujitsu's Digital Annealer development kit to construct the QUBO representation and to ensure that fourth-order terms in \cref{eq:hamiltonian} are effectively reduced to quadratic ones by introducing auxiliary binary variables. QIIO is fully connected, whereas minor embedding is required to map the problem onto QA. After solving the optimization problem \cref{eq:hamiltonian}, the obtained bitstring $\vec{x}$ is used to update $\vec{\mu}$ and $\vec{\omega}$ in accordance with \cref{eq:muomega-increment}, which ultimately yields the next state \(s_{t+1}\). 

The reduction yields $16$ primary variables (four per bus; those of the \emph{slack} bus are overwritten by its set-point) and $120$ (PyQUBO, default penalty strength $5$) or $90$ (QIIO) auxiliary variables. QA used automatic minor embedding, chain strength $2$, and $2000$ reads, of which the lowest-energy one was used. QIIO used $16$ runs of $2000$ iterations with automatic scaling. The bitstring, the unchanged state, and $2000$ random bit-flip perturbations of the bitstring were evaluated by the mean squared mismatch at non-\emph{slack} buses, and the best of these $2000$ candidates formed $s_{t+1}$. The implemented Hamiltonians deviate from \cref{eq:hamiltonian}: the sum included the \emph{slack} bus with zero demand, which penalizes its injection, and for QIIO, the increments of bus $k$ in terms of bus $i$ were scaled by $\Delta\mu_i$, $\Delta\omega_i$ instead of $\Delta\mu_k$, $\Delta\omega_k$. Note that annealers are heuristics without a guarantee of reaching the minimum. The share of the Ising machine in the selected updates relative to the random candidates was not isolated, and solver and quantum processing unit (QPU) times were not recorded.

\begin{figure}[t]
\centering
\includegraphics[width=\columnwidth]{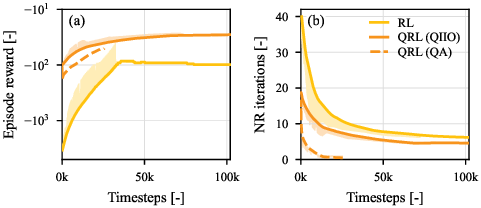}
\caption{Training trajectories of the RL, QRL (QIIO), and QRL (QA) agents for the 4-bus system (QRL variant of the MDP, \cref{sec:mdp}) on common axes: (a) episode reward and (b) number of NR iterations. Lines show values, and shaded areas show the deviation of the raw logged values.}
\label{fig:training}
\end{figure}

\Cref{fig:training} compares the training trajectories of the QRL agents with the RL agent on common axes. Due to the cost of quantum hardware access, the QA experiment is restricted to $\num{2.5e4}$ timesteps. Within $\num{1e5}$ timesteps, the RL agent reduces the number of NR iterations from $40$ to about $6$ and QRL (QIIO) from about $19$ to about $5$, whereas QRL (QA) reaches less than one NR iteration on average within $\num{2.5e4}$ timesteps. Both QRL agents attain higher episode rewards than the RL agent.

\begin{figure}[t]
\centering
\includegraphics[width=\columnwidth]{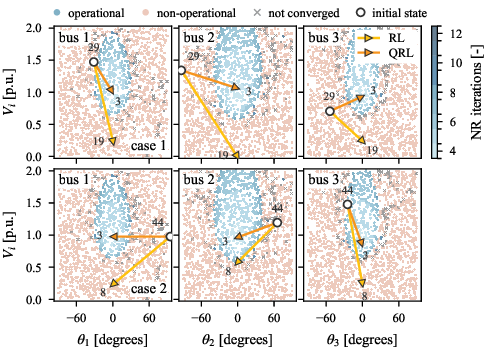}
\caption{Two challenging initial conditions for the 4-bus system (QRL variant of the MDP, \cref{sec:mdp}) on the basin of attraction of \cref{fig:distribution}: voltage magnitude and angle of each \emph{load} bus at the initial state and arrows to the final states of the RL agent (after $20$ timesteps) and of QRL (QIIO) (after one timestep). Numbers: NR iterations from the respective state.}
\label{fig:cases}
\end{figure}

\Cref{fig:cases} presents two challenging initial conditions for the 4-bus system that require $29$ and $44$ NR iterations, respectively. In these cases, QRL reached an initialization requiring three NR iterations within a single RL timestep, whereas the RL agent required $20$ timesteps and reached $19$ and $8$ iterations. These examples illustrate the potential of quantum annealing-based environment updates, which search all combinations of increments for a voltage adjustment with a small PF mismatch in a single step. However, a QRL timestep comprises $2000$ readouts, whereas an RL timestep applies one action. A comparison by RL timesteps, therefore, does not reflect the computational effort.

\subsection{Reactive power limits and PV$\to$PQ switching} \label{sec:qlim}
PV$\to$PQ switching is examined on the 118-bus system (the smaller systems have no PV buses) with $200$ test scenarios at normal loading and at $\lambda \in [0.90, 0.97]\,\hat{\lambda}_Q$, with $11.2$ and $29.3$ of $53$ generators at a limit on average. RL-refined is trained without limits and evaluated without retraining. All runs succeeded, all solutions comply with the limits, and all starts yield the same final generator status and last NR pass ($2.75$ and $2.87$ iterations), so the initialization only affects the first NR pass, as shown in \cref{tab:qlim}. There, RL-refined requires the fewest NR iterations.

\begin{table}[t]
\centering
\caption{118-bus system with generator reactive power limits (test sets, $200$ scenarios per level). Success rate and NR iterations of the first NR pass before PV$\to$PQ switching (RL-refined: mean $\pm$ std over three seeds). The DC start requires one additional (linear) DC PF iteration, which is not included.}
\label{tab:qlim}
\footnotesize
\setlength{\tabcolsep}{4pt}
\begin{tabular}{lccc}
\toprule
 & & \multicolumn{2}{c}{First NR pass: NR iterations} \\
\cmidrule(lr){3-4}
Initialization & Success & Normal & Stressed \\
\midrule
Flat start & 100\% & 4.00 & 5.00 \\
DC start & 100\% & 3.01 & 4.00 \\
RL-refined & 100\% & 2.04 $\pm$ 0.04 & 2.94 $\pm$ 0.10 \\
\bottomrule
\end{tabular}
\end{table}

\section{Discussion}

When it comes to advanced computational technologies, two distinct perspectives emerge. One group of researchers and practitioners advocates their transformative potential, while another remains skeptical of their reliability and real-world applicability. In this study, we present an example that integrates several emerging techniques, i.e., QC, ML, and optimization, to address a long-standing challenge in numerical analysis. Interestingly, the final solution is ultimately obtained by a well-established and trusted classical solver, thus bridging the gap between innovative and conservative approaches, but also making this study appealing to both communities.

Our findings show that RL reduces the NR iteration count beyond flat and DC starts. On all three systems, most of the gain over the DC start stems from the supervised component, and RL contributes a consistent additional reduction across all seeds, even under reactive power limits. A wall-clock gain, however, requires the integration of the initializer into the solver (\cref{sec:cost}). It is well known that the NR method can fail to converge under certain conditions, even when the system of equations is not particularly large. A common cause of such failure is poor initialization of the solver's starting guesses. The basin of attraction of the 4-bus system confirms that an arbitrary initial guess frequently leads to a low-voltage solution. However, flat and DC starts converged in all solvable scenarios considered, so the room for improvement lies in the number of NR iterations. Since NR converges quadratically, a learned initialization must reduce the PF mismatch substantially before an iteration is saved. 

Lessons learned from the experiments performed in this study include (i) the evaluation must verify that the operational solution is reached, and the DC start is the relevant baseline; (ii) the reward must be aligned with the objective, since a small $\|F\|_2$ can also be attained close to a non-operational solution or where the Jacobian is ill-conditioned, whereas the NR iteration count is aligned but integer-valued and saturating; (iii) actions on all bus voltages hamper exploration and destabilize policy updates on large systems, whereas a one-step correction of a supervised initializer~\cite{Okhuegbe2024} in a low-dimensional space was effective; (iv) results vary between seeds, which requires selection on validation data; and (v) under reactive power limits, the initialization only affects the NR pass before PV$\to$PQ switching. 

Among NR's many applications, PF analysis is a critical task in electricity grid planning and operation. The proposed quantum-enhanced RL approach is, however, generalizable to other problems where NR or similar numerical solvers are employed. For example, our previous work has shown that the combinatorial way of modeling existing models can be applied to any system of equations with continuous variables, provided that the objective can be formulated as a root-finding problem~\cite{Kaseb2025solving}.

Nevertheless, the results are subject to several limitations. Stress levels rely on continuation estimates rather than certified loadability limits, and reactive power limits are studied only on the 118-bus system. The conclusions hold for PPO with three seeds per system. The study assumes a fixed topology and bus types. The QRL results are restricted to the 4-bus system by current quantum hardware. Finally, the comparison covers flat and DC starts and one supervised initializer. Other ML-based initializers~\cite{Okhuegbe2024,Chen2020} and convex relaxations~\cite{Molzahn2016} were not evaluated. Future work may linearize the PF mismatch around the current state, which avoids auxiliary QUBO variables, and integrate the learned initializer into the solver.

\section{Conclusion}

We show that NR is sensitive to arbitrary initial points and that the proposed RL-based initialization reduces the number of NR iterations beyond flat and DC starts. Wherever a solution was found on the 4-, 11-, and 118-bus systems, including Iwamoto's ill-conditioned system, stressed loading, and PV$\to$PQ switching, NR from flat and DC starts converged, and stress increased the number of NR iterations rather than causing divergence. RL-refined reduced the NR iterations on all three systems for all seeds, e.g., on the 118-bus system to $2.04$ and $2.86$ under normal and stressed loading, compared with $3.02$ and $5.13$ from DC start and $2.61$ and $3.09$ from the same initializer without RL, also under reactive power limits. The contribution of RL is this incremental gain beyond supervised learning. Since the learned components cost less than one NR iteration, the initializer pays off in wall-clock time if it is integrated into the solver and amortized over many solves on a fixed topology. In the present implementation, PandaPower's built-in DC start remains faster. On the 4-bus system, QRL (QIIO) reached initializations that required three NR iterations within a single RL timestep under challenging initial conditions, illustrating the potential of quantum annealing-based environment updates. Hence, larger systems require a physics-based starting point, a low-dimensional action space, and a reward aligned with the NR iterations. No scalability beyond the 118-bus system or quantum advantage is claimed.

\section*{Acknowledgments}
This work is part of the DATALESs project, with project number 482.20.602, jointly financed by the Netherlands Organization for Scientific Research (NWO) and the National Natural Science Foundation of China (NSFC). This work utilized the Dutch national e-infrastructure, supported by the SURF Cooperative, under grant number EINF-6569. The authors also like to thank Fujitsu Technology Solutions for providing access to the QIIO software\footnote{\url{https://en-portal.research.global.fujitsu.com/kozuchi}} and, in particular, Matthieu Parizy for his support. Furthermore, the authors acknowledge TNO for the access to TNO's Quantum Application Lab Facility, with special thanks to Frank Phillipson for his assistance.

\bibliographystyle{IEEEtran}
\bibliography{references}

\begin{thebibliography}{10}
\providecommand{\url}[1]{#1}
\csname url@samestyle\endcsname
\providecommand{\newblock}{\relax}
\providecommand{\bibinfo}[2]{#2}
\providecommand{\BIBentrySTDinterwordspacing}{\spaceskip=0pt\relax}
\providecommand{\BIBentryALTinterwordstretchfactor}{4}
\providecommand{\BIBentryALTinterwordspacing}{\spaceskip=\fontdimen2\font plus
\BIBentryALTinterwordstretchfactor\fontdimen3\font minus \fontdimen4\font\relax}
\providecommand{\BIBforeignlanguage}[2]{{%
\expandafter\ifx\csname l@#1\endcsname\relax
\typeout{** WARNING: IEEEtran.bst: No hyphenation pattern has been}%
\typeout{** loaded for the language `#1'. Using the pattern for}%
\typeout{** the default language instead.}%
\else
\language=\csname l@#1\endcsname
\fi
#2}}
\providecommand{\BIBdecl}{\relax}
\BIBdecl

\bibitem{Cossent2011}
R.~Cossent, L.~Olmos, T.~G{\'o}mez, C.~Mateo, and P.~Fr{\'i}as, ``Distribution network costs under different penetration levels of distributed generation,'' \emph{European Transactions on Electrical Power}, vol.~21, no.~6, pp. 1869--1888, 2011.

\bibitem{Delabays2022}
R.~Delabays, S.~Jafarpour, and F.~Bullo, ``Multistability and anomalies in oscillator models of lossy power grids,'' \emph{Nature Communications}, vol.~13, no.~1, p. 5238, 2022.

\bibitem{Milano2009}
F.~Milano, ``Continuous {Newton's} method for power flow analysis,'' \emph{IEEE Transactions on Power Systems}, vol.~24, no.~1, pp. 50--57, 2009.

\bibitem{deOliveira2023}
L.~N. de~Oliveira, F.~D. Freitas, and N.~Martins, ``A modal-based initial estimate for the {Newton} solution of ill-conditioned large-scale power flow problems,'' \emph{IEEE Transactions on Power Systems}, vol.~38, no.~5, pp. 4962--4965, 2023.

\bibitem{Irving1987}
M.~R. Irving and M.~J.~H. Sterling, ``Efficient {Newton-Raphson} algorithm for load-flow calculation in transmission and distribution networks,'' \emph{IEE Proceedings C -- Generation, Transmission and Distribution}, vol. 134, no.~5, p. 325, 1987.

\bibitem{Wang2025}
R.~Wang, G.~Zhang, W.~Xiong, B.~Wang, W.~Wei, and M.~Wei, ``Ill-conditioned power flow calculation in urban rail traction power supply system,'' \emph{IEEE Transactions on Transportation Electrification}, vol.~11, no.~3, pp. 8462--8473, 2025.

\bibitem{Alharbi2021}
T.~Alharbi, M.~Tostado-V{\'e}liz, O.~Alrumayh, and F.~Jurado, ``On various high-order {Newton}-like power flow methods for well and ill-conditioned cases,'' \emph{Mathematics}, vol.~9, no.~17, p. 2019, 2021.

\bibitem{Stott2009}
B.~Stott, J.~Jardim, and O.~Alsa{\c{c}}, ``{DC} power flow revisited,'' \emph{IEEE Transactions on Power Systems}, vol.~24, no.~3, pp. 1290--1300, 2009.

\bibitem{AbdelAkher2015}
M.~Abdel-Akher, A.~Selim, and M.~M. Aly, ``Initialised load-flow analysis based on {Lagrange} polynomial approximation for efficient quasi-static time-series simulation,'' \emph{IET Generation, Transmission \& Distribution}, vol.~9, no.~16, pp. 2768--2774, 2015.

\bibitem{Coutinho2023}
C.~Coutinho~de Oliveira, A.~Bonini~Neto, D.~A. Alves, C.~R. Minussi, and C.~A. Castro, ``Alternative current injection {Newton} and fast decoupled power flow,'' \emph{Energies}, vol.~16, no.~6, p. 2548, 2023.

\bibitem{Barati2024}
M.~Barati, ``Enhancing {ACPF} analysis: Integrating {Newton-Raphson} method with gradient descent and computational graphs,'' arXiv:2406.10390, 2024.

\bibitem{Wang2017}
B.~Wang, J.~Bachan, and C.~Chan, ``{ExaGridPF}: A parallel power flow solver for transmission and unbalanced distribution systems,'' in \emph{2018 IEEE Power \& Energy Society Innovative Smart Grid Technologies Conference (ISGT)}, 2018, pp. 1--5.

\bibitem{Trias2012}
A.~Trias, ``The holomorphic embedding load flow method,'' in \emph{2012 IEEE Power and Energy Society General Meeting}, 2012, pp. 1--8.

\bibitem{Okhuegbe2024}
S.~N. Okhuegbe, A.~A. Ademola, and Y.~Liu, ``A machine learning initializer for {Newton-Raphson} {AC} power flow convergence,'' in \emph{2024 IEEE Texas Power and Energy Conference (TPEC)}, 2024, pp. 1--6.

\bibitem{Chen2020}
L.~Chen and J.~E. Tate, ``Hot-starting the {AC} power flow with convolutional neural networks,'' arXiv:2004.09342, 2020.

\bibitem{Molzahn2016}
D.~K. Molzahn and I.~A. Hiskens, ``Convex relaxations of optimal power flow problems: An illustrative example,'' \emph{IEEE Transactions on Circuits and Systems I: Regular Papers}, vol.~63, no.~5, pp. 650--660, 2016.

\bibitem{Kraft1986}
L.~A. Kraft and G.~T. Heydt, ``Adaptive acceleration factors for the {Newton-Raphson} power flow study,'' \emph{Electric Machines \& Power Systems}, vol.~11, no.~4, pp. 337--346, 1986.

\bibitem{Yang2024}
N.-C. Yang and C.-H. Tseng, ``Adaptive convergence enhancement strategies for {Newton-Raphson} power flow solutions in distribution networks,'' \emph{IET Generation, Transmission \& Distribution}, vol.~18, no.~13, pp. 2339--2352, 2024.

\bibitem{Aviles2024}
J.~Avil{\'e}s, D.~Guillen, L.~Ibarra, and J.~D. D{\'a}valos-Soto, ``Reconfiguration of active distribution networks as a means to address generation and consumption dynamic variability,'' \emph{IET Generation, Transmission \& Distribution}, vol.~18, no.~19, pp. 3120--3137, 2024.

\bibitem{Zimmerman2011}
R.~D. Zimmerman, C.~E. Murillo-S{\'a}nchez, and R.~J. Thomas, ``{MATPOWER}: Steady-state operations, planning, and analysis tools for power systems research and education,'' \emph{IEEE Transactions on Power Systems}, vol.~26, no.~1, pp. 12--19, 2011.

\bibitem{Zeng2022}
L.~Zeng and H.-D. Chiang, ``Mulitple power flow solutions due to {PV-PQ} switching and their properties,'' in \emph{2022 IEEE Power \& Energy Society General Meeting (PESGM)}, 2022, pp. 1--5.

\bibitem{Song2024}
Y.~Song, T.~Liu, and Y.~Hou, ``Voltage stability constrained optimal power flow considering {PV-PQ} bus type switching: Formulation and convexification,'' \emph{IEEE Transactions on Power Systems}, vol.~39, no.~2, pp. 3336--3348, 2024.

\bibitem{Sutton2018}
R.~S. Sutton and A.~G. Barto, \emph{Reinforcement Learning: An Introduction}, 2nd~ed.\hskip 1em plus 0.5em minus 0.4em\relax Cambridge, MA: MIT Press, 2018.

\bibitem{Ma2024}
C.~Ma, A.~Li, Y.~Du, H.~Dong, and Y.~Yang, ``Efficient and scalable reinforcement learning for large-scale network control,'' \emph{Nature Machine Intelligence}, 2024.

\bibitem{Wang2025b}
Y.~Wang, X.~Yu, and W.~Zhang, ``An improved reinforcement learning-based differential evolution algorithm for combined economic and emission dispatch problems,'' \emph{Engineering Applications of Artificial Intelligence}, vol. 140, p. 109709, 2025.

\bibitem{Yang2020}
Q.~Yang, G.~Wang, A.~Sadeghi, G.~B. Giannakis, and J.~Sun, ``Two-timescale voltage control in distribution grids using deep reinforcement learning,'' \emph{IEEE Transactions on Smart Grid}, vol.~11, no.~3, pp. 2313--2323, 2020.

\bibitem{Chicano2025}
F.~Chicano, G.~Luque, Z.~A. Dahi, and R.~Gil-Merino, ``Combinatorial optimization with quantum computers,'' \emph{Engineering Optimization}, vol.~57, no.~1, pp. 208--233, 2025.

\bibitem{Morstyn2024}
T.~Morstyn and X.~Wang, ``Opportunities for quantum computing within net-zero power system optimization,'' \emph{Joule}, vol.~8, no.~6, pp. 1619--1640, 2024.

\bibitem{Aharonov2007}
D.~Aharonov, W.~van Dam, J.~Kempe, Z.~Landau, S.~Lloyd, and O.~Regev, ``Adiabatic quantum computation is equivalent to standard quantum computation,'' \emph{SIAM Journal on Computing}, vol.~37, no.~1, pp. 166--194, 2007.

\bibitem{Barends2016}
R.~Barends, A.~Shabani, L.~Lamata, J.~Kelly, A.~Mezzacapo, U.~Las~Heras, R.~Babbush, A.~G. Fowler, B.~Campbell, Y.~Chen \emph{et~al.}, ``Digitized adiabatic quantum computing with a superconducting circuit,'' \emph{Nature}, vol. 534, no. 7606, pp. 222--226, 2016.

\bibitem{Kaseb2025solving}
Z.~Kaseb, M.~M{\"o}ller, P.~Palensky, and P.~P. Vergara, ``A framework for solving continuous energy and power system problems using adiabatic quantum computing,'' arXiv:2504.06458, 2025.

\bibitem{Iwamoto1981}
S.~Iwamoto and Y.~Tamura, ``A load flow calculation method for ill-conditioned power systems,'' \emph{IEEE Transactions on Power Apparatus and Systems}, vol. PAS-100, no.~4, pp. 1736--1743, 1981.

\bibitem{Tinney1967}
W.~Tinney and C.~Hart, ``Power flow solution by {Newton's} method,'' \emph{IEEE Transactions on Power Apparatus and Systems}, vol. PAS-86, no.~11, pp. 1449--1460, 1967.

\bibitem{Thurner2018}
L.~Thurner, A.~Scheidler, F.~Schafer, J.-H. Menke, J.~Dollichon, F.~Meier, S.~Meinecke, and M.~Braun, ``pandapower---an open-source {Python} tool for convenient modeling, analysis, and optimization of electric power systems,'' \emph{IEEE Transactions on Power Systems}, vol.~33, no.~6, pp. 6510--6521, 2018.

\bibitem{Ajjarapu1992}
V.~Ajjarapu and C.~Christy, ``The continuation power flow: a tool for steady state voltage stability analysis,'' \emph{IEEE Transactions on Power Systems}, vol.~7, no.~1, pp. 416--423, 1992.

\bibitem{Konda1999}
V.~R. Konda and J.~N. Tsitsiklis, ``Actor-critic algorithms,'' in \emph{Advances in Neural Information Processing Systems}, vol.~12.\hskip 1em plus 0.5em minus 0.4em\relax MIT Press, 1999.

\bibitem{Schulman2017}
J.~Schulman, F.~Wolski, P.~Dhariwal, A.~Radford, and O.~Klimov, ``Proximal policy optimization algorithms,'' arXiv:1707.06347, 2017.

\bibitem{Raffin2021}
A.~Raffin, A.~Hill, A.~Gleave, A.~Kanervisto, M.~Ernestus, and N.~Dormann, ``{Stable-Baselines3}: Reliable reinforcement learning implementations,'' \emph{Journal of Machine Learning Research}, vol.~22, no. 268, pp. 1--8, 2021.

\bibitem{Grainger1994}
J.~J. Grainger and W.~D. Stevenson, Jr., \emph{Power System Analysis}.\hskip 1em plus 0.5em minus 0.4em\relax New York: McGraw-Hill, 1994.

\bibitem{kaseb2024power}
Z.~Kaseb, M.~M{\"o}ller, P.~P. Vergara, and P.~Palensky, ``Power flow analysis using quantum and digital annealers: a discrete combinatorial optimization approach,'' \emph{Scientific Reports}, vol.~14, no.~1, p. 23216, 2024.

\end{thebibliography}

\end{document}